\documentclass[12pt]{article}
\usepackage{epsfig}
\usepackage{latexsym}
\usepackage{axodraw}

\begin{document}

\begin{titlepage}

\baselineskip 24pt

\begin{center}

{\Large {\bf Implications of a Rotating Mass Matrix}}\\

\vspace{.5cm}

\baselineskip 14pt

{\large Jos\'e BORDES}\\
jose.m.bordes\,@\,uv.es\\
{\it Departament Fisica Teorica, Universitat de Valencia,\\
  calle Dr. Moliner 50, E-46100 Burjassot (Valencia), Spain}\\

\vspace{.2cm}

{\large CHAN Hong-Mo}\\
chanhm\,@\,v2.rl.ac.uk  \,\,\,  \\
{\it Rutherford Appleton Laboratory,\\
  Chilton, Didcot, Oxon, OX11 0QX, United Kingdom}\\

\vspace{.2cm}

{\large TSOU Sheung Tsun}\\
tsou\,@\,maths.ox.ac.uk\\
{\it Mathematical Institute, University of Oxford,\\
  24-29 St. Giles', Oxford, OX1 3LB, United Kingdom}

\end{center}

\vspace{.3cm}

\begin{abstract}

The fermion mass matrix, in addition to having eigenvalues (masses) which
run, also changes its orientation (rotates) with changing energy scales.
This means that its eigenstates at one scale will no longer be eigenstates
at another scale, leading to effects where fermions of different flavours
can ``transmute'' into one another.  In this paper, the implications of
a rotating mass matrix are analysed and possible transmuation effects are 
investigated both in the Standard Model (SM) and in the so-called Dualized 
Standard Model (DSM) that we advocate, arriving at the conclusion that 
some transmutational decays such as $\psi \longrightarrow \mu \tau$, 
$\Upsilon \longrightarrow \mu \tau$ or $\pi^0 \longrightarrow e \mu$ may 
be within experimental range, if not immediately, then in the near future.

\end{abstract}

\end{titlepage}

\clearpage

\baselineskip 14pt

\setcounter{section}{0}
\setcounter{equation}{0}
\def\theequation{\arabic{section}.\arabic{equation}}

\section{Introduction - the Rotating Mass Matrix}

By a rotating mass matrix, we mean one which undergoes unitary transformations
through scale changes.  

As most quantities in quantum field theory vary under changing scales, in 
particular the masses of particles or in other words the eigenvalues of mass
matrices, it is no surprise that the orientations of mass matrices may also
change.  

Indeed, even in the Standard Model as usually conceived, nontrivial mixing 
between up and down fermion states will generally induce rotating mass 
matrices.  That this is so can be seen from the following renormalization 
group equations:
\begin{equation}
16 \pi^2 \frac{dU}{dt} = \frac{3}{2} (UU^\dag - DD^\dag)U
   +(\Sigma_u - A_u)U,
\label{rgeq1}
\end{equation}
\begin{equation}
16 \pi^2 \frac{dD}{dt} = \frac{3}{2} (DD^\dag - UU^\dag)D
   +(\Sigma_d - A_d)D.
\label{rgeq2}
\end{equation}
satisfied by the mass matrices $U$ and $D$ of respectively the up and down 
fermions \cite{rge}.  Nontrivial mixing between up- and down-states 
means that the 
matrices $U$ and $D$ are related by a nondiagonal matrix $D = VUV^\dag$ 
or $U = V^\dag DV$, so that $U$ and $D$ cannot be simultaneously digonalized. 
Suppose now at some $t$ we diagonalize $U$ in (\ref{rgeq1}) by an appropriate 
unitary transformation.  All other terms in (\ref{rgeq1}) are then diagonal but
not the term $DD^\dag U$.  Hence, this term will necessarily de-diagonalize
the matrix $U$ on running by (\ref{rgeq1}) to a different $t$, or in other 
words the matrix $U$ will rotate with changing $t$ \cite{ramond}.  
Similar arguments hold 
also for the rotation of the matrix $D$.  Given that quarks have long been 
known to have a nontrivial (CKM) mixing matrix \cite{ckm}, and that 
recent experiments 
on neutrino oscillation \cite{Soudan,chooz,superk} strongly suggest a 
nontrivial 
(MNS) mixing \cite{mns} also 
for leptons, we have to conclude that the mass matrices for both quarks 
and leptons will rotate with changing scales even if there are no forces 
and interactions in nature other than those currently studied in the 
Standard Model.

However, there is clearly a possibility---perhaps one might even say a 
strong theoretical reason to suppose---that forces may exist in nature 
other than those now conventionally studied in the Standard Model, which
can give further rotations to the fermion mass matrices.  Indeed, in 
the usual formulation of the Standard Model, the fact that there are 
3 generations of fermions and that they mix is taken as an input from 
experiment, with the result that the framework depends on a large number 
of empirical parameters, about three-quarters of which are traceable to
the mystery of generations.  One could thus hope that in future when 
the generation puzzle is solved, some or even most of the empirical 
parameters appearing in the present Standard Model will be predictable.  
One favourite and perhaps most natural assumption for theoretical attempts 
in this direction is that generations originate as a broken `horizontal' 
gauge symmetry \cite{horizontal}.  If that is the case, then the 
interactions associated 
with this symmetry which mix the generation index are likely also to 
rotate the fermion mass matrices in generation space.  This rotation would 
be over and above that driven by the nondiagonal CKM or MNS matrices via 
the mechanism described in the preceding paragraph.

Whatever its origin, however, the rotation of fermion mass matrices is 
theoretically of the same standing as the running of masses and coupling 
constants, and if testable by experiment would lend equal support to the 
quantum field theory framework.  Moreover, if experiment shows that the 
rotation is different from simply that given by renormalization group
equations (\ref{rgeq1}) and (\ref{rgeq2}) via the empirical mixing matrix,
then it suggests that there are forces at work other than those currently
studied in the Standard Model which are of the type considered in the
preceding paragraph.  This means that we would have opened up a new window
for investigations furthering our long-standing quest for the origin of 
fermion generations.  And since, as we shall see, mass matrix rotation 
leads to an entirely new and very distinctive category of physical phenomena, 
there is a fair chance of its effects being observed by experiment in the
not too distant future.  The purpose of the present paper is to explore 
such possibilities.

\setcounter{equation}{0}

\section{Specification of Fermion States}

The fact that the mass matrix rotates poses immediately a question of 
physical interpetation at a rather basic level, namely the question of how 
to define the state vectors of the various fermion states, and hence the 
mixing matrices between them.  For the familiar case of a non-rotating 
mass matrix, one defines the state vectors of the 3 generations as its 
eigenvectors, which are by assumption scale-independent and, since the 
matrix is hermitian, are also mutually orthogonal as they should be if 
they are to represent independent physical entities.  The mixing matrix
between the up and down fermion types, being the overlap matrix between 
2 scale-independent orthonormal triads of state vectors, one for each 
type, is then also scale-independent and automatically unitary.  Only 
the eigenvalues of a mass matrix then run, and the masses of the 3 
states (generations) can be defined as the running values each evaluated 
at the scale equal to its value.  However, when the mass matrix rotates, 
although one can still diagonalize the matrix at any scale by a triad 
of mutually orthogonal eigenvectors, this triad will be scale-dependent,
and it is not immediately clear which vectors at which scale(s) are to be 
identified as the state vectors of the 3 generations.  Although the ambiguity
may not in all cases be numerically significant in view of present limitations
in experimental accuracy, it still has to be resolved as a matter of principle
against the day when better accuracy is achieved.   

Now, the state vector of a physical state like the muon is normally, we 
think, taken as a scale-independent concept, or otherwise we are likely to 
meet with some awkwardness in its physical interpretation.  For example, 
suppose we were to define the muon state as the eigenvector with the second 
highest eigenvalue of the charged lepton mass matrix at any scale.  Then 
since the matrix rotates, the $\mu$ state vector will point at different 
scales in different directions in generation space.  In that case, the 
muons obtained say from an energetic beam of charged $\pi$'s, which we 
know decay almost entirely into $\mu$'s at rest, will no longer appear as 
purely muons when it hits the target, but as a linear combination of 
$e, \mu$ and $\tau$, which is at variance with what is usually understood.  
Although one can in principle insist on defining fermion states as the 
eigenstates of the rotating mass matrix at every scale, and so long as 
one is consistent in their interpretation one would arrive in the end 
at the same physical results, such a procedure would seem to be rather 
inconvenient.  In this paper therefore, we opt for a scale-independent
definition of fermion state vectors, in which case it would be incumbent 
upon us to specify exactly the scale at which each of these vectors are to 
be defined.

If the rotation of the mass matrix is negligible to a certain approximation 
below some scale, then one may define to this approximation the state 
vectors as the orthonormal triad of eigenvetors of the mass matrix taken 
at that scale.  This, though seldom stated explicitly is, we think, the 
tacit criterion usually adopted in the literature which we shall refer 
to in this paper as Fixed Scale Diagonalization FSD.  It is, of course, 
at best an approximation for the mass matrix cannot stop rotating 
abruptly, and that approximation, as we shall see later, may not be all 
that good in certain circumstances.  Besides, this criterion involves a
degree of arbitrariness in choosing a certain fixed scale to effect the 
diagonalisation of the mass matrix, and begs the question of principle
why it should be that particular scale and not some other. 

In view of this, one may be tempted instead to do the following.  As one 
defines the mass of a state as the eigenvalue evaluated at the scale equal 
to its value, one may try to define as its state vector the corresponding 
eigenvector also at the same scale.  This would be more democratic and less 
arbitray than the FSD criterion but it will not work.  Consider for example 
the $U$-type quarks.  Following the above proposal, one would then define 
the $t$ state vector as the eigenvector with the largest eigenvalue $m_3$ 
of the matrix evaluated at the scale $\mu = m_3(\mu)$, the $c$ state vector
as the eigenvector with the second largest eigenvalue $m_2$ at the scale
$\mu = m_2(\mu)$, and the $u$ state vector as the eigenvector with the
smallest eigenvalue $m_1$ at the scale $\mu = m_1(\mu)$.  These 3 state
vectors, however, will not be mutually orthogonal, for although the 3 
eigenvectors of the hermitian mass matrix are mutually orthogonal when all 
evaluated at the same scale, they have no reason to be so when evaluated 
each at a vastly different scale since the matrix by assumption rotates with
changing scales.  It thus contradicts the assertion that they represent 
3 independent physical entities; in particular, it would imply that the 
mixing (CKM or MNS) matrix, being a transformation matrix between the 2 
triads of up and down state vectors, is not unitary as it ought to be.

There is, however, a working criterion for defining state vectors each at 
its own mass scale along the lines suggested in the above paragraph which 
takes account of the rotation between the different mass scales yet still
gives mutually orthogonal state vectors and hence a unitary mixing matrix.
This was first proposed by us \cite{dualcons} in connection with a scheme
for quark and lepton mixing which we call the Dualized Standard Model (DSM),
but the criterion can in fact be applied to define state vectors in any 
scheme with a rotating mass matrix \cite{phenodsm}.  For 3 generations, the 
criterion goes as follow.  We run the mass matrix $m$ down in scale until 
we have for its highest eigenvalue $m_3$ a solution to the equation 
$\mu = m_3(\mu)$.  This value at this scale we define as the mass $m_3$, 
and the corresponding eigenvector the state vector ${\bf v_3}$ of the
heaviest generation.  Below that energy, the state 3 no longer exists as 
a physical state, and only the two lighter generations survive, whose
state vectors have to be orthogonal to ${\bf v_3}$.  We define then
the mass matrix at energies below $m_3$ as the $2 \times 2$ submatrix 
$\hat{m}$ of $m$ in the subspace orthogonal to ${\bf v_3}$.  To find
now the mass and state vector for generation 2, we follow with $\hat{m}$
the same procedure used above with $m$ for generation 3 and run $\hat{m}$
down in scale until we find a solution to the equation $\mu = \hat{m}_2(\mu)$,
which value we call the mass $m_2$ and the corresponding eigenvector at
that scale the state vector ${\bf v_2}$ of the generation 2.  The state
vector of the lightest generation 1 is now also defined as the vector
orthogonal to both ${\bf v_3}$ and ${\bf v_2}$, while the mass of 1 is 
obtained by repeating the same procedure, namely by running down in scale
the remaining $1 \times 1$ submatrix, namely the expectation value
$\langle {\bf v_1}|m|{\bf v_1} \rangle$, until its value equals the scale.
In this way, each mass is evaluated at its own appropriate scale while
the physical state vectors of the 3 generations are mutually orthogonal
as they should be.  Furthermore, the mixing matrix, taken as the overlap
matrix between the triads of state vectors so defined for the up and
down states, will be both unitary and scale-independent.  This criterion
for defining state vectors, which is the only one we are aware of with
the required properties, we shall refer to in future as Step-by-Step
Diagonalization SSD.

\setcounter{equation}{0}

\section{Fermion Transmutation as Consequence}

Whether one accepts the SSD proposal for defining physical fermion state 
vectors or chooses to ignore these subtleties and adopts the approximate 
criterion of FSD, the fact remains that the mass matrix will continue to 
rotate at high energy, and so long as the state vectors are defined as 
the eigenvectors diagonalizing the mass matrix at some prescribed scale(s), 
they will generally no longer diagonalize the mass matrix at some higher 
energy scales.  Now, the fermion mass matrix appears for example in
the fermion 
propagator of Feynman diagrams, and if it is nondiagonal, so also will 
be the transition amplitudes represented by these diagrams.  Physically,
this means that transitions will be induced in which a particle of one 
generation converts into a particle of another generation.  For example,
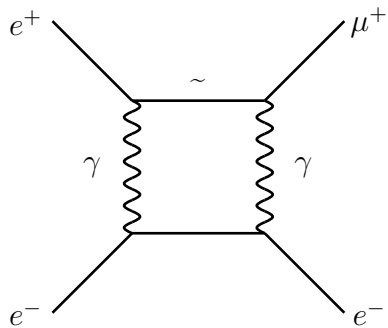
\begin{figure}[ht]
\vskip 2cm
\begin{center}
{\unitlength=1.0 pt \SetScale{1.0} \SetWidth{1.0}
\begin{picture}(150,100)(0,0) 
\Line(50,50)(100,50)
\Photon(50,50)(50,100){3}{6}
\Photon(100,50)(100,100){3}{6}
\Line(50,100)(100,100)

\Line(50,50)(20,20)
\Line(100,100)(130,130)
\Line(100,50)(130,20)
\Line(50,100)(20,130)

\Text(35,75)[]{$\gamma$}
\Text(115,75)[]{$\gamma$}
\Text(75,105)[]{$\widetilde{}$}
\Text(10,130)[]{$e^+$}
\Text(140,130)[]{$\mu^+$}
\Text(10,20)[]{$e^-$}
\Text(140,20)[]{$e^-$}

\end{picture} }
\end{center}
\caption{Transmutation of $e$ into $\mu$ in $e^+ e^-$ collision.}
\label{eptomup}
\end{figure}
the diagram of Figure \ref{eptomup}, where a tilde denotes the occurrence 
of a nondiagonal fermion mass matrix, will have an off-diagonal element 
linking the $e$ state vector to the $\mu$ state vector, leading to a
$e^+ e^- \longrightarrow \mu^+ e^-$ transition.

Such a conversion of $e$ to $\mu$ is quite distinctive and singular,  
differing from that induced, for example, by a flavour-changing neutral 
current (FCNC) exchanged between the initial $e^+$ and $e^-$.  The latter
will normally produce in addition to the $\mu^+$ at the upper vertex also 
a $\mu^-$ at the lower vertex, giving rise to a ``double conversion'', 
although under certain circumstances where the exchanged FCNC bosons 
themselves mix, ``single conversion'' can also occur, as for example 
in the situation considered in \cite{mueconv}.  However, even in that 
last case, the conversion rate depends on the masses and couplings 
of the exchanged FCNC bosons which are still rather elusive quantitites, 
whereas the conversion effect considered here is due only to the rotation 
of the mass matrix, and may therefore be quite restricted.  We propose 
therefore to call the latter ``transmutation'' to distinguish it from 
other conversion phenomena.

The effect of transmutation is so unique as a consequence of the rotating 
mass matrix that if it occurs with appreciable cross sections, it should
be identifiable in experiment with little difficulty.  At first sight,
therefore, the prediction looks alarming, for the conservation of such 
quantities as muon number have been checked already to high accuracy, and 
their wholesale violation as suggested above seems hardly likely to survive 
the existing experimental bounds.  On closer examination, however, the
conclusion becomes less obvious, mainly for the following reason.  At
low energy, near the scale(s) at which the mass matrix is diagonalized 
to define the physical fermion states, as explained in the preceding 
section, the deviations of the mass matrix from diagonality are small,
and so also will be the cross sections for transmutation.  At high energy,
on the other hand, where the off-diagonal mass matrix elements can
become sizeable, it turns out that transmutation cross sections will be 
suppressed, as can be seen in the following example.

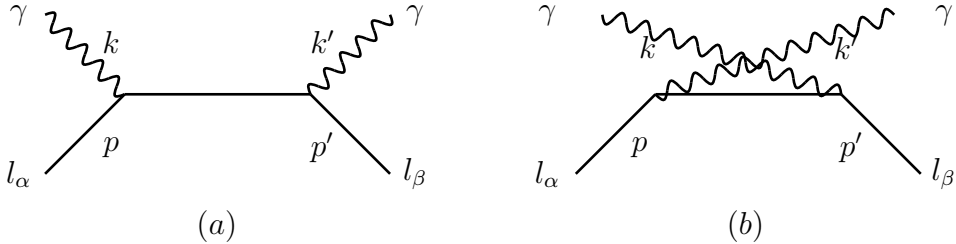
\begin{figure}[ht]
%\vskip 2cm
\begin{center}
{\unitlength=1.0 pt \SetScale{1.0} \SetWidth{1.0}
\begin{picture}(350,100)(0,0) 
\Line(30,50)(100,50)
\Photon(30,50)(0,80){3}{5}
\Photon(100,50)(130,80){3}{5}
\Line(0,20)(30,50)
\Line(100,50)(130,20)

\Text(-10,20)[]{$l_\alpha$}
\Text(140,20)[]{$l_\beta$}
\Text(-10,80)[]{$\gamma$}
\Text(140,80)[]{$\gamma$}
\Text(25,30)[]{$p$}
\Text(105,30)[]{$p'$}
\Text(25,70)[]{$k$}
\Text(105,70)[]{$k'$}
\Text(65,0)[]{$(a)$}

\Line(230,50)(300,50)
\Photon(230,50)(320,80){-3}{10}
\Photon(300,50)(210,80){-3}{10}
\Line(200,20)(230,50)
\Line(300,50)(330,20)
\Text(190,20)[]{$l_\alpha$}
\Text(340,20)[]{$l_\beta$}
\Text(190,80)[]{$\gamma$}
\Text(340,80)[]{$\gamma$}
\Text(225,30)[]{$p$}
\Text(305,30)[]{$p'$}
\Text(228,68)[]{$k$}
\Text(303,68)[]{$k'$}
\Text(265,0)[]{$(b)$}
\end{picture} }
\end{center}
\caption{Feynman diagrams for photo-transmutation of leptons.}
\label{Comptdiag}
\end{figure}

One of the simplest transmutation processes that one can think of is what 
one can call the photo-transmutation of leptons, which to leading order 
is given just by the Feynman diagrams in Figure \ref{Comptdiag} 
which are the same as for ordinary Compton scattering except that the 
fermion line here carries a rotating and generally non-diagonal mass 
matrix.  Consider first the $s$-channel diagram (a) in which the lepton
propagator $1/(p\llap/ + k\llap/ - m)$ appears, carrying the non-diagonal 
mass matrix $m$ where transmutation originates.  For $s = (p + k)^2$ 
large, we can expand the propagator in powers of $m/\sqrt{s}$, obtaining
symbolically:
\begin{equation}
\frac{1}{(p\llap/ + k\llap/ - m)} \sim \frac{1}{(p\llap/ + k\llap/)}
   (1 + \frac{m}{\sqrt{s}}),
\label{leppropa}
\end{equation}
where one sees that the off-diagonal terms leading to transmutation is
suppressed by a factor of $1/\sqrt{s}$.  Similarly, the lepton propagator
in the diagram (b) for $u$ large can be approximated as:
\begin{equation}
\frac{1}{(p\llap/ - k\llap/' - m)} \sim \frac{1}{(p\llap/ - k\llap/')}
   (1 + \frac{m}{\sqrt{|u|}}),
\label{leppropb}
\end{equation}
leading to transmutation effects suppressed by a factor $1/\sqrt{|u|}$.  In
the region where $|u|$ is small, off-diagonal contributions can remain 
sizeable, but the size of this region itself decreases as $1/s$ so that
the integrated effect remains small.  

For these reasons, transmutation effects arising from the rotating mass 
matrix are seldom very large and the question whether they can already 
be ruled out by existing data, and if not, how soon, and under what 
conditions, they will be verifiable, is not immediately obvious and can 
only be answered by detailed analysis.  The answers will depend on the 
process investigated and on the speed with which the mass matrices rotate, 
and these in turn depend on the theory or model one is considering.  We 
examine below and in a companion paper \cite{photrans} some specific 
examples.

\setcounter{equation}{0}

\section{Two Rotation Mechanisms}

In this section, we investigate the rotating fermion mass matrix in 2 
specific schemes at, as it were, opposite ends of the spectrum so as 
hopefully to span most reasonable possibilities.  At one end of the 
spectrum, we assume that there are no forces in nature other than those
currently studied in the Standard Model so that all the rotation in the 
fermion mass matrix is driven by the electroweak Higgs terms via the 
empirical nondiagonal mixing (CKM or MNS) matrix.  The generation 
phenomenon is thereby left unexplained.  Fermion state vectors are to
be specified by the FSD prescription (to contrast with the scheme below
which only works with the SSD prescription) in which the mass matrix is
diagonalized at some chosen (low) scale.  In the case of leptons, since 
some mass and mixing parameters are but poorly known, these specifications 
have yet to be supplemented by further assumptions based on current popular 
views in a manner to be detailed later.  We shall refer to this henceforth
as the NSM scheme (for naively implemented Standard Model).   At the 
other end of the spectrum, we consider a scheme we ourselves suggested 
\cite{phenodsm}, which in turn was abstracted from our so-called Dualized 
Standard Model (DSM) \cite{dualcons,ourCKM} although it need not strictly 
adhere to the duality concepts of that earlier work.  In this scheme, 
generation itself originates from a broken $SU(3)$ gauge symmetry, the 
rotation is driven by Higgs bosons associated with the breaking of this 
symmetry, and even the nondiagonal mixing matrices as experimentally 
observed are themselves consequences of the mass matrix rotation.  Here 
the SSD prescription for defining state vectors is essential.  Given that 
in the DSM it is the rotation which gives rise to fermion mixing, and not 
the other way round as in the NSM case with the mixing driving the 
rotation, transmutation effects are usually, though not in all cases, 
larger in the DSM than in the NSM, and hence would be easier either to 
disprove or to confirm.      

Our first task is to evaluate the rotating mass matrix over a range of 
scales, say, up to the energy achievable in the foreseeable future of around 
100 TeV.  For the NSM case, this in principle requires the solution of 
the equations (\ref{rgeq1}) and (\ref{rgeq2}) together with any other 
renormalization group equations they are coupled to, which though standard 
is a little involved.  For our present purpose of a first exploration, 
however, we may approximate by linearizing the equations and keeping only 
those terms actually driving the rotation, namely the term $D D^\dag U$ 
in (\ref{rgeq1}) and the term $U U^\dag D$ in (\ref{rgeq2}).  We need 
as input from experiment the masses of the fermion states and the mixing 
matrices measured at low energies.  Assuming that the scale in the FSD 
prescription chosen for defining the fermion state vectors coincides with
the scale at which the empirical mixing matrices are measured, the rotating 
mass matrices can then be evaluated by iterating the two equations.  For 
quarks, both the masses and the CKM matrix have now been determined to 
sufficient accuracy for our purpose and pose no practical difficulty.
This calculation has been performed but since it will not be of use in 
what follows in this paper, the result will not be presented.  For 
leptons, however, the MNS matrix is still poorly known, while the Dirac 
masses of neutrinos on which their Higgs couplings depend are still almost 
entirely unconstrained.  For these, therefore, we shall just insert for 
our exploration the most popular theoretical biases at present, namely 
for the MNS matrix the so-called bi-maximal mixing \cite{bimax} version:
\begin{equation}
U_{MNS} = \left( \begin{array}{ccc} 1/\sqrt{2} & 1/\sqrt{2} & 0 \\
                                           1/2 & 1/2 & 1/\sqrt{2} \\
                                           1/2 & 1/2 & 1/\sqrt{2}
                                           \end{array} \right), 
\label{bimaxckm}
\end{equation}
and for the Dirac mass $m_3$ of the heaviest neutrino a value around the mass 
of the $t$ quark.  One shall need the Dirac mass $m_2$ of the second heaviest 
neutrino also, which is taken, for lack of any better choice, to be of
the order of the charm mass.  Since the equations have been linearized 
in approximation, the result of the rotation for charged leptons can be 
summarized simply by giving the rates of change of the off-diagonal
elements of the (`hermitized') mass matrix with the logarithm of the 
energy, which are explicitly 
given as:
\begin{equation}
\frac{d L}{d t} = \frac{3}{128 \pi^2} \frac{1}{V^2} 
   \left( \begin{array}{ccc} 
   2 m_2^2 m_e & \frac{1}{\sqrt{2}} m_2^2 (m_e + m_\mu) & 
      \frac{1}{\sqrt{2}} m_2^2 (m_e + m_\tau) \\
   \frac{1}{\sqrt{2}} m_2^2 (m_e + m_\mu) & 2 m_3^2 m_\mu &
      m_3^2 (m_\mu + m_\tau) \\
   \frac{1}{\sqrt{2}} m_2^2 (m_e + m_\tau) & m_3^2 (m_\mu + m_\tau) &
      2 m_3^2 m_\tau \end{array} \right),
\label{rorate}
\end{equation}   
with $V$ being the vev of the electroweak Higgs, i.e.\ 246 GeV.  Putting 
in then the suggested values of the masses, one obtains that $\langle \mu 
| m | \tau \rangle$ changes by about $5.5 \times 10^{-3}$, $\langle e 
| m | \tau \rangle$ by about $1.77 \times 10^{-7}$ and $\langle e | m | 
\mu \rangle$ by about $1.06 \times 10^{-8}$ GeV per decade change in energy.  
These results will be useful later for estimating the rates of various 
transmutation effects implied by the NSM scheme.  One notices that in 
(\ref{rorate}), $\langle e | m | \tau \rangle$ is proportional to $m_2^2$ 
only by virtue of the assumed zero in the bi-maximal mixing matrix.
If that element $U_{e3}$ of the MNS matrix is nonzero, then $\langle e 
| m | \tau \rangle$ would be proportional to $m_3^2$, which in turn can
lead on iteration to terms of order $m_3^4$ in the element $\langle e 
| m | \mu \rangle$, thus making both much larger than the above estimates, 
which represent therefore a sort of lower limit.

\begin{figure}
\centering
\includegraphics{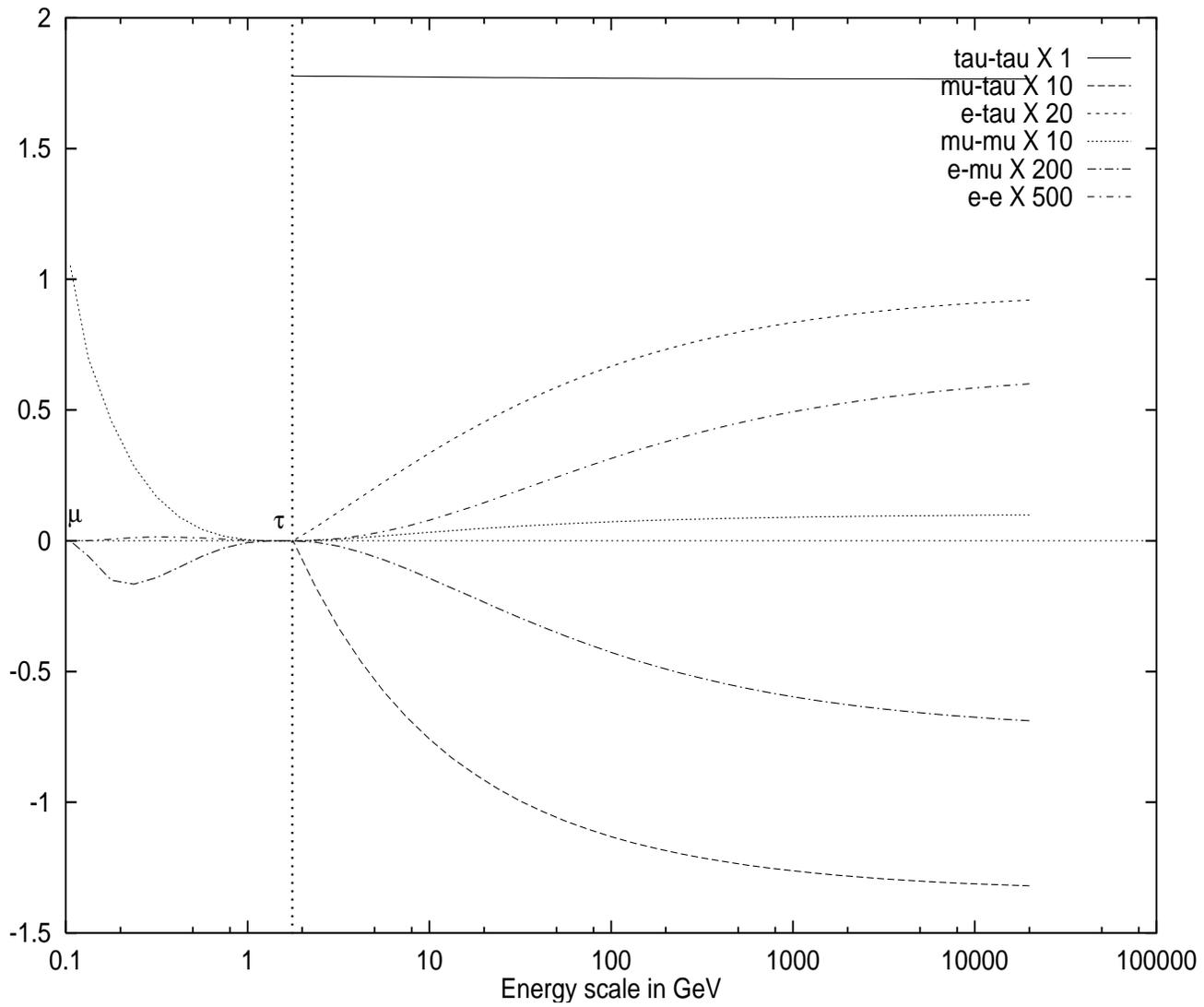}
\caption{Elements of the rotating mass matrix in GeV for charged 
leptons in the DSM scheme} 
\label{leptonm}
\end{figure}

For the DSM case, the calculation, though more intricate, has been more 
explicitly specified.  As the scheme depends on only 3 parameters 
\cite{phenodsm} which have already been determined by fitting to some 
and then used to give very sensible predictions to others of the mass
and mixing parameters of both quarks and leptons, no further information
needs to be supplied.  Indeed, in making these fits and predictions, the
rotating mass matrices for quarks and leptons had perforce already been
evaluated, though not previously presented.  The result for the charged 
leptons which is of concern to us in this paper is now given in Figures 
\ref{leptonm}.  

To check our result, without going through the details of the fit in 
\cite{phenodsm}, one needs only to iterate the following equation for a 
vector $(x',y',z')$:
\begin{equation}
\frac{d}{dt} \left( \begin{array}{c} x' \\ y' \\ z' \end{array} \right)
   = \frac{5}{32 \pi^2} \rho^2 \left( \begin{array}{c} x_1' \\ y_1' \\ z_1'
      \end{array} \right),
\label{dsmroeq}
\end{equation}
with
\begin{equation}
x_1' = \frac{x'(x'^2-y'^2)}{x'^2+y'^2} + \frac{x'(x'^2-z'^2)}{x'^2+z'^2},
   \ \ \ {\rm cyclic},
\label{x1prime}
\end{equation}
starting from some initial value $(x_I, y_I, z_I)$ with a coupling $\rho$, 
all given in eq. 16 of \cite{phenodsm}.  For an accuracy of about a percent,
some 500 iterations are needed per decade change in energy, where the vector
$(x', y', z')$ is to be normalized after every iteration.  The resulting 
mass matrix is of the factorized form:
\begin{equation}
m' = m_T \left( \begin{array}{c} x' \\ y' \\ z' \end{array} \right)
      (x', y', z'),
\label{massmat}
\end{equation}
where $m_T$ is a normalization depending on the fermion-type $T$, which
may be taken as the mass of the heaviest generation state of that type.
In (\ref{massmat}), the mass matrix is given with respect to a fixed 
(gauge) basis in generation space, but the matrix with respect to the 
physical basis as presented in, for example, Figure \ref{leptonm} is 
also easily evaluated with the (physical) state vectors determined in 
accordance with the SSD prescription outlined in section 2.  Thus, the 
physical state vectors of the charged leptons in the gauge basis are,
explicitly:
\begin{eqnarray}
{\bf v}_\tau & = & (0.9967, 0.0760, 0.0268) \nonumber \\ 
{\bf v}_\mu  & = & (-0.0759, 0.7741, 0.6285) \nonumber \\
{\bf v}_e    & = & (0.0271, -0.6285, 0.7774).
\label{lepvectors}
\end{eqnarray}
For more details, the reader is referred to \cite{phenodsm} and earlier
references therein. 

One particular feature of the DSM scheme is that all fermions lie on the 
same trajectory of the rotating vector $(x',y',z')$ so that at any scale the 
mass matrix for up and down fermion types have the same eigenvectors.  This 
means that the $U$ and $D$ matrices in (\ref{rgeq1}) and (\ref{rgeq2}) can 
be simultaneously diagonalized, causing thus no de-diagonalization to each 
other via the electroweak Higgs loop effect.  We recall that it was the 
latter which was what drove the rotation in the NSM case.  Here 
instead all the rotation comes from the Higgs bosons associated with the 
broken `horizontal' symmetry.  

Comparing the actual values for the rotating mass matrices of the two 
schemes NSM and DSM, one finds the off-diagonal elements to be generally
larger for the latter than the former, as we expected.  Further, as can be 
seen in Figure \ref{leptonm}, the mass matrix in the DSM scheme has quite 
an intricate structure with, for example, the off-diagonal elements 
passing through zero whenever the energy scale equals the mass of a 
physical fermion state, meaning that the mass matrix is there diagonal.
This is a consequence of the SSD prescription by construction, and would
apply also to the NSM had one chosen there also the SSD instead of the
FSD prescription.

\setcounter{equation}{0}

\section{Examples of Transmutational Decays}

Once given the rotating mass matrix as a function of the energy scale, 
it would appear that one has in principle already enough information 
to evaluate any implied transmutational effect.  However, because of 
the mass matrix rotation leading to, among other things, some rather 
unfamiliar kinematics, adaptations to usual calculational procedures are 
required, which are not, at first encounter, entirely straight-forward. 
We find that this is true even in evaluating, for example, the simple 
Feynman diagrams of Figure \ref{Comptdiag} for the photo-transmutation of 
leptons that we have calculated.  Therefore, in order to avoid clouding
immediately the essentially simple general picture of transmutation with 
technical details, we relegate detailed calculations to other papers, e.g.
on photo-transmutation to a companion paper \cite{photrans}, and limit 
ourselves here to qualitative estimates of widths for some illustrative 
examples in transmutational decays, which being single particle effects 
are simpler to analyse theoretically.  Also, to avoid the complications
of quark confinement, we shall deal here with only lepton transmutations,
which are already of considerable physical interest, being constrained 
by some very stringent bounds in experiment.

(I) As a first example, consider the decays:
\begin{equation}
Z^0 \longrightarrow l_\alpha {\bar l}_\beta, 
\label{Zdecays}
\end{equation}
with $l_\alpha \neq l_\beta$ being different lepton states.  As indicated 
already in Section 3, the amplitudes of such off-diagonal transmutational 
processes are expected to be suppressed with respect to those of their 
diagonal counterparts $Z^0 \longrightarrow l_\alpha {\bar l}_\alpha$ by a 
factor of the form $\langle \alpha|m|\beta \rangle/E$, with $E$ being a
measure of the typical energy carried by the fermion line, which we can
take here to be of the order of (half) the $Z^0$ mass $M_Z$.  The numerator
$\langle \alpha|m|\beta \rangle$ represents the off-diagonal mass matrix
element linking the lepton states $l_\alpha$ and $l_\beta$ at the scale
of the reaction, which is again $M_Z$ in the present case.  Given any
choice for $l_\alpha$ and $l_\beta$, $\langle \alpha|m|\beta \rangle$ can
be read off for the DSM in Figure \ref{leptonm}.  Thus, for example, for
the decay $Z^0 \longrightarrow \tau^- \mu^+$ the relevant matrix element
$\langle \mu |m|\tau \rangle$ at the scale $M_Z = 91$GeV is seen to have
value about $-0.11$ GeV.  For the NSM, the element $\langle \mu |m|\tau 
\rangle$ was said in Section 3 to vary by about 0.5 percent per decade change
in energy.  According to the FSD prescription adopted there, we have to
first fix a scale at which the mass matrix is diagonal.  If this is taken,
for lack of any obvious better choice, to be at the $\tau$ mass, we obtain
for $\langle \mu |m|\tau \rangle$ at $M_Z$ a value of about 0.01 GeV.
To estimate the width for the decay $Z^0 \longrightarrow \tau^- \mu^+$,
the easiest way would be to compare it with either of the diagonal decays
$Z^0 \longrightarrow \tau^- \tau^+$ or $Z^0 \longrightarrow \mu^- \mu^+$.
Given that the masses of the final state leptons are so much smaller than 
the decaying $Z^0$ mass, one can neglect the small differences in kinematics
between the different decay modes and write simply:
\begin{equation}
\frac{\Gamma(Z^0 \longrightarrow \tau^- \mu^+)}
   {\Gamma(Z^0 \longrightarrow \tau^- \tau^+)}
   \sim \frac{|\langle \mu |m|\tau \rangle|^2}{M_Z^2}.
\label{Ztaumu}
\end{equation}
Putting in the above estimates of the matrix elements and the empirical 
value 3.36 percent \cite{databook} for the branching ratio of the diagonal
$\tau^- \tau^+$ mode, one obtains then for the branching ratio of
the transmutational mode $\tau^- \mu^+$ an estimate of about $4 \times
10^{-8}$ for DSM and $4 \times 10^{-10}$ for NSM.  Both of these estimates
are much below the present empirical bound of $1.2 \times 10^{-5}$ listed 
in \cite{databook}.  Similarly, estimates for the branching ratios of
the other modes are obtained, giving for the $\tau^- e^+$ mode a 
branching ratio of $4 \times 19^{-9}$ and for the $\mu^- e^+$ mode
$1.6 \times 10^{-11}$ for DSM.  Again these are way below the present
empirical limits listed in \cite{databook} of respectively $9.8 \times 
10^{-6}$ and $1.7 \times 10^{-6}$.  The estimates in the NSM for these 
last two modes are minuscule due to the very small values for the mass
matrix elements linking $e$ to $\mu$ and $e$ to $\tau$, as will also be
true for the following examples.  They shall therefore henceforth be 
ignored except, for a particular reason to be explained, for (IV) below.
As one can easily see, the reason that the transmutation rates are all so 
small in $Z^0$ decay is that the off-diagonal effects are much suppressed 
by the high energies involved which is here of the order of the $Z^0$ mass.  

(II) As a second example then, let us go to the other extreme in 
energy and consider the decay:
\begin{equation}
\pi^0 \longrightarrow \mu^- e^+
\label{pitomue}
\end{equation}
which in $\pi^0$-decay is the only kinematically accessible transmutational 
leptonic mode.  The same arguments as before suggests that we compare this
with the diagonal mode $\pi^0 \longrightarrow e^- e^+$ giving a suppression
in the amplitude of the former with respect to the latter by a factor of
the form $\langle e|m|\mu \rangle/m_\pi$ where the off-diagonal mass matrix
element is to be evaulated at the scale $m_\pi$.  Here, the spins, parities
and masses of the particles involved being such, a correction for the 
difference in kinematics between the two decays is warranted, which is 
made most easily by comparison with the analogous charged $\pi$ decays, 
giving thus:
\begin{equation}
\frac{\Gamma(\pi^0 \longrightarrow \mu^- e^+)}
   {\Gamma(\pi^0 \longrightarrow e^- e^+)} 
   \sim \frac{|\langle e|m|\mu \rangle|^2}{m_\pi^2}
   \frac{\Gamma(\pi^\pm \longrightarrow \mu^\pm \nu_\mu)}
        {\Gamma(\pi^\pm \longrightarrow e^\pm \nu_e)}.
\label{pimue}
\end{equation}
The matrix element $\langle e|m|\mu \rangle$ at scale $m_{\pi^0} = 135$ MeV 
obtained from the calculation giving Figure \ref{leptonm} is about
$3 \times 10^{-4}$ GeV.  Hence, putting in the empirical value of 
$7.5 \times 10^{-8}$ for the branching ratio of the diagonal mode 
$\pi^0 \longrightarrow e^- e^+$ and that of $1.23 \times 10^{-4}$ for
the branching ratio of $\Gamma(\pi^\pm \longrightarrow e^\pm \nu_e)$
\cite{databook}, we have an estimate for the branching ratio of the 
transmuational mode $\pi^0 \longrightarrow \mu^- e^+$  of about 
$2.9 \times 10^{-9}$.  This is barely an order of magnitude below the 
experimental bound of $1.72 \times 10^{-8}$ \cite{databook}.  Indeed,
one is saved here from a violation of the experimental limit only by 
the exceptionally small value of the off-diagonal mass matrix element 
due to the proximity of the reaction scale $m_{\pi^0}$ to the muon 
mass where the mass matrix is constrained to be diagonal by the SSD 
prescription.

(III) As a third example, let us consider at an intermediate 
scale:
\begin{equation}
\psi \longrightarrow l_\alpha {\bar l}_\beta.
\label{psitoll}
\end{equation}
Following the same procedure as above, one obtains:
\begin{equation}
\frac{\Gamma(\psi \longrightarrow \tau^- \mu^+)}
   {\Gamma(\psi \longrightarrow \mu^- \mu^+)}
   \sim \frac{|\langle \mu |m|\tau \rangle|^2}{m_\psi^2}.
\label{psitaumu}
\end{equation}
The matrix element $\langle \mu |m|\tau \rangle$ for DSM as read from Figure
\ref{leptonm} at the scale $m_\psi$ is about $0.03$ GeV.  Combined with the
empirical value of about 6 percent \cite{databook} for the branching ratio 
of $\psi \longrightarrow \mu^- \mu^+$, one obtains an estimate of about
$6 \times 10^{-6}$ for the transmutational $\tau^- \mu^+$ mode, which is
larger than in either of the two earlier examples.  The corresponding 
estimate for the NSM is about $1 \times 10^{-8}$.  Although no empirical
limit is given in \cite{databook} for this particular mode, the estimates
seem close enough to the sensitivity of present experiment to make a 
search for it worthwhile.  Similar arguments applied to the modes 
$\psi \longrightarrow \tau^- e^+$ and $\psi \longrightarrow \mu^- e^+$ 
yield the estimated branching ratios $1.6 \times 10^{-7}$ and $6 \times 
10^{-11}$.  For these modes again, no empirical limits are given in 
\cite{databook} but the estimates seem in any case beyond present experimental
sensitivity.  The same anlysis can also be applied to $\Upsilon$ decay
giving for the branching ratios of the modes $\tau^- \mu^+, \tau^- e^+$,
and $\mu^- e^+$ respectively the following estimates: $1.6 \times 10^{-6},
\ 6 \times 10^{-8}$, and $7 \times 10^{-11}$.  Again, no empirical bounds 
for these modes are given in \cite{databook}, but some, especially the 
$\tau \mu$ mode, will be worth searching for in future in e.g. $B$-factories.

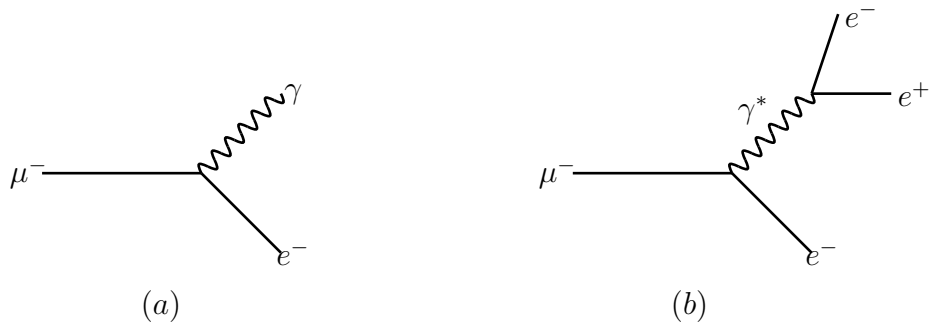
\begin{figure}[ht]
%\vskip 2cm
\begin{center}
{\unitlength=1.0 pt \SetScale{1.0} \SetWidth{1.0}
\begin{picture}(350,100)(0,0)
\Line(0,30)(60,30)
\Photon(60,30)(90,60){3}{6}
\Line(60,30)(90,0)

\Text(-5,30)[]{$\mu^-$}
\Text(95,60)[]{$\gamma$}
\Text(95,0)[]{$e^-$}
\Text(45,-20)[]{$(a)$}

\Line(200,30)(260,30)
\Photon(260,30)(290,60){3}{6}
\Line(260,30)(290,0)
\Line(290,60)(300,90)
\Line(290,60)(320,60)
\Text(195,30)[]{$\mu^-$}
\Text(269,54)[]{$\gamma^*$}
\Text(310,90)[]{$e^-$}
\Text(330,60)[]{$e^+$}
\Text(295,0)[]{$e^-$}
\Text(245,-20)[]{$(b)$}
\end{picture} }
\end{center}
\vskip 0.5cm
\caption{Transmutational $\mu^- \longrightarrow e^- \gamma$ and 
   $\mu^- \longrightarrow e^- e^+ e^-$ decays}
\label{muegamma}
\end{figure}

(IV) Lastly, consider transmutational fermion decay, of which 
the prime examples would be:
\begin{equation}
\mu^- \longrightarrow e^- \gamma
\label{mutoeg}
\end{equation}
and
\begin{equation}
\mu^- \longrightarrow e^- e^+ e^-
\label{mutoeee}
\end{equation}
proceeding via the Feynman diagrams in Figure \ref{muegamma}.
The bound as given in \cite{databook} for the branching ratio of the
decay (\ref{mutoeg}) is an impressive $4.9 \times 10^{-11}$, and of the
decay (\ref{mutoeee}) is $1 \times 10^{-12}$.  And these coming on top 
of a main 
decay mode of $\mu$, namely $\mu \longrightarrow e \nu {\bar \nu}$, proceeding 
by weak interactions with a width of only around $3 \times 10^{-10}$ eV, are 
a very stringent constraint indeed on any scheme which violates muon number.
It is thus gratifying for us to note that transmutation in the DSM scheme 
survives this very dangerous-looking hurdle automatically.  This comes 
about because the Feynman diagrams \ref{muegamma} for the transmutational
decays (\ref{mutoeg}) and (\ref{mutoeee}) are to be evaluated at the scale 
of the $\mu$ mass, and at that scale the mass matrix in the DSM scheme is 
diagonal.  That it is so by virtue of the SSD prescription for defining
lepton states was noted already in preceding sections and can be seen 
explicitly in Figure \ref{leptonm} by the vanishing of the off-diagonal 
mass matrix element $\langle e|m| \mu \rangle$ at the $\mu$ mass.  Hence 
the DSM scheme actually predicts zero probability for this transmutational 
decay.   

The same observation applies also to transmutational decays of $\tau$, for 
example:
\begin{equation}
\tau^- \longrightarrow e^- \gamma,\ \mu^- \gamma,
\label{tautolg}
\end{equation}
\begin{equation}
\tau^- \longrightarrow e^- e^+ e^-,\ e^- \mu^+ \mu^-,\ \mu^- e^+ e^-,
   \ \mu^- \mu^+ \mu^-,
\label{tautolll}
\end{equation}
since for these decays the analogous Feynman diagrams are to be evaluated 
at the scale of the $\tau$ mass, and at that scale the mass matrix in the 
DSM scheme is again diagonal giving thus zero transitions.  This is also 
fortunate, for although not quite as stringent as those on (\ref{mutoeg})
and (\ref{mutoeee}), the bounds on the branching ratios of (\ref{tautolg}) 
and (\ref{tautolll}) as given in \cite{databook} are still only of order 
$10^{-6}$ for a total $\tau$ width of around $3 \times 10^{-3}$eV, which 
would be far from trivial to satisfy otherwise.

\begin{figure}[ht]
\begin{center}
{\unitlength=1.0 pt \SetScale{1.0} \SetWidth{1.0}
\begin{picture}(150,100)(0,0)
\Line(0,30)(60,30)
\Photon(60,30)(90,60){3}{6}
\Line(60,30)(90,0)
\Line(90,60)(100,90)
\Line(90,60)(120,60)
\Text(-5,30)[]{$\mu^-$}
\Text(70,55)[]{$W$}
\Text(110,90)[]{$e^-$}
\Text(130,60)[]{$\bar{\nu}_e$}
\Text(95,0)[]{$\nu_\mu$}
\end{picture} }
\end{center}
\caption{Feynman diagram for the decay $\mu^- \longrightarrow e^- 
{\bar \nu}_e \nu_\mu$}
\label{muenunu}
\end{figure}
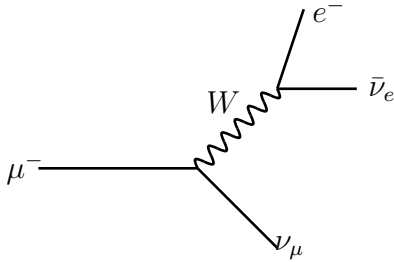

To illustrate how sensitive a test these lepton decays are to transmutation 
models, let us examine them in the NSM scenario.  Here, as noted before, 
the rotation of the mass matrix is slower than in the DSM case giving 
thus usually smaller transmutation rates, especially for transmutations
between $\mu$ and $e$.  However, the mass matrix is required by the FSD 
prescription adopted to be diagonal only at some fixed mass scale.  Suppose 
we choose this scale as the $\tau$ mass so as to make the transmutation 
rates vanish for both the decays (\ref{tautolg}) and (\ref{tautolll}) and 
so satisfy the experimental bounds on them automatically.  As the energy 
lowers to the scale of the $\mu$ mass, however, the mass matrix will have 
rotated and acquired nonzero off-diagonal elements.  Indeed, according 
to (\ref{rorate}) the element $\langle e | m | \mu \rangle$ would have 
acquired on running from the $\tau$ mass a value of around $1.3 \times 
10^{-8}$ GeV, giving thus possibly nonzero rates to the decays (\ref{mutoeg}) 
and (\ref{mutoeee}).  To estimates these rates following the previous 
procedure, let us compare for example (\ref{mutoeee}) to the known main 
decay mode of $\mu$, namely:
\begin{equation}
\mu^- \longrightarrow e^- {\bar \nu}_e \nu_\mu,
\label{mutoenunu}
\end{equation}
as given to first order by the Feynman diagram of Figure \ref{muenunu}.
The mass of the electron being so much smaller than that of the muon,
we can ignore the small difference in phase space between the two decays 
and write:
\begin{equation}
\frac{\Gamma(\mu^- \longrightarrow e^- e^+ e^-)}
   {\Gamma(\mu^- \longrightarrow e^- {\bar \nu}_e \nu_\mu)}
   = \frac{|\langle e | m | \mu \rangle|^2}{m_\mu^2} \frac{e^2/m_\mu^2}
     {g_2^2/M_W^2}
\label{BRmueee}
\end{equation}
where we have taken for a typical momentum for the decay to be of order
of the muon mass.  Inserting the above NSM value for $\langle e | m | 
\mu \rangle$ and the empirical values for the other quantities appearing 
in (\ref{BRmueee}), one obtains a value of about $1 \times 10^{-3}$ for 
the ratio.  This would be disastrous, being some 9 orders too large 
compared with the experimental bound given in \cite{databook} of order 
$1 \times 10^{-12}$.  That this ratio turns out to be so large despite 
the very small NSM value for $\langle e | m | \mu \rangle$ is because of 
the large $W$ mass occuring in the diagram for the normal decay mode of 
$\mu$, or in other words, because we are in (\ref{BRmueee}) comparing what 
is basically, in spite of being off-diagonal, still a photon-mediated 
process to a weak decay.  The above blatant violation of the experimental 
bound is of course not the fault of the Standard Model but only of the
naive fashion that it has been implemented in the NSM scheme since it 
could be easily avoided by, among other means, choosing a smaller value
for the Dirac mass $m_2$ of the second heaviest neutrino which is still
at present entirely unconstrained by experiment.  Nevertheless, this 
example shows how sensitive these lepton decays are as a test of 
transmutation models, and how delicately the DSM scheme has managed to 
survive it.

Obviously, what really mattered for passing that test was what we called
the SSD prescription for defining the lepton state vectors which stipulates
that the mass matrix is to be diagonal at the mass of every lepton state.
Had one chosen in the NSM scheme to define fermion states in this way 
instead of the FSD prescription adopted, it would have passed the test also.
With the FSD prescription, however, it would not have mattered if one had 
chosen the scale for diagonalizing the mass matrix to be $m_\mu$ instead 
of $m_\tau$, for although one would then pass the test for the decay
$\mu^- \longrightarrow e^- e^+ e^-$, one would fail it for such decays as
$\tau^- \longrightarrow \mu^- \mu^+ \mu^-$.  Indeed, running now from the
$\mu$ mass, one would obtain a non-diagonal mass matrix at the $\tau$ mass
with off-diagonal elements of the same order.  The empirical bounds on
transmutational $\tau$ decays are less stringent, e.g.\ the present bound
on the branching ratio of $\tau^- \longrightarrow \mu^- \mu^+ \mu^-$ is
only $1.9 \times 10^{-6}$ \cite{databook}, but that would still mean a
violation of the bound by some 6 orders in magnitude by the above estimate.
One could of course impose in the NSM a further condition on the FSD 
prescription stating that the mass matrix should stop rotating altogether 
once below the chosen diagonalization scale, but one does not know of a 
good theoretical reason for doing so.  

Firm conclusions cannot be drawn until detailed calculations for the 
various decay widths have been performed, which to us seem quite feasible
in the near future.  Tentatively, however, one can claim that the DSM
predictions for transmutation survive present experimental bounds in
all cases so far studied, while the NSM (with FSD) survives except perhaps
for the noted $\mu$ or $\tau$ transmutational decays.  In most cases, the
predicted rates are some way below present experimental limits, but for
the decays $\pi^0 \longrightarrow e^\pm \mu^\mp$, $\psi \longrightarrow
\mu^\pm \tau^\mp$ and $\Upsilon \longrightarrow \mu^\pm \tau^\mp$ the
estimated rates may soon be, if they are not already, within range of 
experimental sensitivity.

\setcounter{equation}{0}

\section{Remarks}

The importance of possible fermion flavour violation has long been
recognized and subjected to rigorous test by such experiments with
ultra-high sensitivity as $\mu \longrightarrow e \gamma$ decay.  The
negative outcome to-date of all these tests conveys the impression that 
such violations can exist, if at all, only to a very small extent.  However,
that conclusion was reached in the days before it was realized that the
fermion mass matrix actually rotates with the energy scale so that the
fact that flavour is conserved at one scale, or that the mass matrix is 
there diagonal in the flavour states, does not necessarily mean that it 
will remain so at some other scales.  That being the case, the accuracy, 
though impressive, of such experiments as $\mu \longrightarrow e \gamma$ 
searches, is not enough to rule out flavour violation in general but has 
to be supplemented by experiments at other energies.  As our analysis 
above has shown, under some circumstances the mass matrix can acquire 
quite appreciable off-diagonal elements without being noticed in existing 
experiments.  Hence, regardless of which scheme for mass matrix rotation 
one favours, it would be worthwhile for experimenters to perform routine 
checks under varying circumstances for the diagonality of fermion mass 
matrices by looking for effects of the type labelled in this paper as 
transmutations.  With a rotating mass matrix, diagonality is no longer
as sacrosanct as once believed.

As argued in the Introduction, that the mass matrix rotates forces on us a 
refinement of our definition of fermion flavour states.  Given the sensitivity 
of such experiments as $\mu^- \longrightarrow e^- e^+ e^-$ searches, 
even the small rotations which may occur between the mass scales of two
successive generations is enough to give huge violations of the experimental 
limits, as we have shown in the preceding section.  It seems thus necessary
either to define the fermion states in such a way as to ensure that the mass
matrix is diagonal at the mass scales of all the fermion states, as was 
done in the SSD prescription, or else in a prescription like FSD to
suppress altogether the rotations between fermion states, if not by some
theoretical mechanism then by decree.  For us, this conclusion has come as
an agreeable surprise in that the SSD prescription which was originally
invented just for internal consistency of our DSM scheme should turn out now
to have the unexpected virtue of guaranteeing consistency with experiment
on $\mu \longrightarrow e \gamma$ and $\mu^- \longrightarrow e^- e^+ e^-$
decays, which would otherwise be a difficult hurdle to survive.  As we 
shall see in \cite{photrans}, the same property of the SSD prescription 
will save one also from certain unwanted pole structures in the amplitudes 
for the photo-transmutation of leptons.

The rotation of its mass matrix being weaker, the NSM scheme will give
generally smaller transmutation effects than the DSM, i.e.\ provided that it 
has been amended with something like the SSD prescription for defining fermion
states to avoid the noted probable inconsistencies.  Indeed, in all the 
examples so far studied NSM transmutation effects seem to be beyond present
experimental sensitivity.  That being the case, if in the routine check 
on transmutation bounds an experiment comes up with an effect, then it 
is probably due to a rotation of the mass matrix driven by some forces 
of nature other than those currently considered in the Standard Model.  
This would be exciting in any case, besides possibly opening up a new window 
for probing into the generation mystery, as we have already said in the 
Introduction.

For us advocates of the Dualized Standard Model (DSM), the results of this 
paper and its companion \cite{photrans} are something of a relief, at least 
temporarily.  Since the whole DSM scheme relies heavily on the concept 
of a rotating mass matrix driven by dual colour forces, on which all its 
seemingly sensible predictions on the mass and mixing patterns of fermions
also depend, it would be a little disastrous if the relatively fast
mass matrix rotation required for the scheme's success predicts at
the same time much larger transmutation effects than can be accommodated
by present experiment.  And, in contrast to flavour-changing neutral
current (FCNC) effects \cite{airshowers,fcnc,mueconv} which also pose
some hurdles for the scheme to overcome but depend on an unknown mass
parameter, there is in transmutations no free parameter which could
be adjusted as a loophole for escape.  It is thus fortunate that a
disaster does not seem so far to have happened although in some cases
it has come rather dangerously close.  Instead, one obtains as
a bonus a whole new class of phenonmena, with quantitative predictions
which, while remaining below present bounds, are still not that far off
as to be entirely inaccessible to experiment.  However, much more work
will still be needed first, to make sure that disasters will not happen
in circumstances not yet examined, and secondly, to identify specific
cases where the scheme's predictions can be checked by experiment, if
not immediately, at least in the not too distant future.

In general terms, what we have done in this paper is to raise some
basic questions of concept forced on us by the rotation of the fermion
mass matrix and attempt to answer some of them.  Although in answering
these questions we may have raised as many more, we hope at least to
have clarified a little some of their implications.

\end{document}